# Dipolar Hole-Blocking Layers for Inverted Perovskite Solar Cells: Effects of Aggregation and Electron Transport Levels

Julian F. Butscher[1,2], Qing Sun[1,2], Yufeng Wu[3], Fabian Stuck[3], Marvin Hoffmann[4], Andreas Dreuw[5], Fabian Paulus[1,2], A. Stephen K. Hashmi[3,5], Nir Tessler[6] and Yana Vaynzof[1,2]*

[1] Kirchhoff Institute for Physics, Heidelberg University, Im Neuenheimer Feld 227, 69120 Heidelberg, Germany
[2] 2 Integrated Centre for Applied Physics and Photonic Materials and Centre for Advancing Electronics Dresden (cfaed), Technical University of Dresden, Nöthnitzer Straße 61, 01187 Dresden, Germany
[3] Organic Chemistry Institute, Im Neuenheimer Feld 270, Heidelberg University, 69120 Heidelberg, Germany.
[4] Interdisciplinary Center for Scientific Computing (IWR), Heidelberg University, Im Neuenheimer Feld 205A, 69120 Heidelberg
[5] Chemistry Department, Faculty of Science, King Abdulaziz University (KAU), 21589 Jeddah (Saudi Arabia)
[6] Sara and Moshe Zisapel Nano-Electronic Center, Department of Electrical Engineering, Technion-Israel Institute of Technology, Haifa 32000, Israel

*E-mail: yana.vaynzof@tu-dresden.de

## Abstract

Herein, we report on the synthesis and investigation of two triazino-isoquinoline tetrafluoroborate electrolytes as hole-blocking layers in methylammonium triiodide perovskite photovoltaic devices with fullerene electron extraction layer. We find that increasing the thickness of the dipolar hole-blocking layer results in a gradual increase in the open-circuit voltage suggesting that aggregation of the molecules can enhance the dipole induced by the layer. This finding is confirmed by theoretical calculations demonstrating that while both molecules exhibit a similar dipole moment in their isolated state, this dipole is significantly enhanced when they aggregate. Ultra-violet photoemission spectroscopy measurements show that both derivatives exhibit a high ionisation potential of 7 eV, in agreement with their effective hole-blocking nature demonstrated by the devices. However, each of the molecules shows a different electron affinity due to the increased conjugation of one of the derivatives. While the change in electron transport level between the two derivatives is as high as 0.3 eV, the difference in the open-circuit voltage of both types of devices is negligible, suggesting that the electron transport level plays only a minor role in determining the open-circuit voltage of the device. Numerical device simulations confirm that the increase in built-in potential, arising from the high dipole of the electrolyte layer, compensates for the non-ideal energetic alignment of the charge transport levels, resulting in high VOC for a range of electron transport levels. Our study demonstrates that the application of small molecule electrolytes as hole-blocking layer in inverted architecture perovskite solar cells is a powerful tool to enhance the open-circuit voltage of the devices and provides useful guidelines for designing future generations of such compounds.

Keywords: Open-circuit voltage; Perovskite solar cells; Small molecule electrolytes; Aggregation; Device simulation.

# 1. Introduction

Since their discovery, perovskite solar cells (PSCs) have had an incredible journey of increasing popularity and performance, now reaching over 25 % power conversion efficiency.[1] While planar PSCs can be fabricated in two architectures, defined by the order of transport layers in the device structure, the inverted or p-i-n architecture in particular received significant attention due to its ease of fabrication[2,3] and suppressed hysteresis[4,5]. Such devices consist of a bottom hole transport layer (HTL) and a top electron transport layer (ETL) that sandwich the perovskite active layer. A range of materials have been investigated as HTLs, including inorganic oxides[6–9], polymers[10–12] or small molecules[13–15] yielding varying photovoltaic performances. Among these materials, one of the most extensively studied polymer HTL for perovskite solar cells is Poly(3,4-ethylenedioxythiophene)-poly(styrenesulfonate) (PEDOT:PSS). While this HTL is easy to process and has been applied in a range of photovoltaic devices[16–18], its low work-function, that leads to an unfavourable energy mismatch at the PEDOT:PSS/perovskite interface, significantly reduces the performance of devices employing PEDOT:PSS as HTL. In particular, the open-circuit voltage ($V_{OC}$) suffers as compared to devices using other higher work-function HTLs as e.g. poly[bis(4-phenyl) (2,4,6-trimethylphenyl)amine] (PTAA)[19–21] or $NiO_x$[22–24].

On the electron extracting side, polymers[25], or more commonly, fullerenes such as $C_{60}$ or [6,6]-phenyl-$C_{61}$-butyric acid methyl ester (PCBM) are employed as it has been shown that they minimise the J-V hysteresis by trap state passivation[26]. However, the relatively low ionization potential of fullerenes as compared to the valence band of the perovskite active layer results in an undesirable hole conduction towards the cathode, causing significant loss in fill factor. To eliminate such losses, an additional layer of a material with a high ionization potential is typically added at the fullerene/cathode interface and is termed a 'hole-blocking layer'. The positive effect on the fill factor of devices has been shown for many HBLs, including metal acetylacetonate[27], LiF[28], titanium (diisopropoxide) bis(2,4-pentanedionate)[29], and perylene−diimide[30], yet one of most commonly employed HBLs remains bathocuproine (BCP) due to its ease of fabrication and relative stability[31–33]. Most HBLs have little to no effect on the $V_{OC}$ of the devices, especially in the case of PEDOT:PSS as a HTL. Nevertheless, we recently showed that a sufficiently large dipole moment within the BCP HBL can remove all limitations to the $V_{OC}$ set by the lower work function of PEDOT:PSS compared to other HTLs[36]. Other approaches to overcome the unfavourable energetics of PEDOT:PSS compared to other HTLs focus on improving the properties of PEDOT:PSS by doping[18,34] and interface engineering[35].

However, only few reports investigate a combined approach in which a HBL is effective in both hole-blocking and enhancing the built-in potential of the device to compensate for the unfavourable energetics on the HTL side, resulting in the simultaneous increase in both fill factor and $V_{OC}$ [37,38].

In this work, we report on the synthesis and study of two triazino-isoquinoline tetrafluoroborate (TI-$BF_4$) electrolytes as HBL in inverted architecture methylammonium triiodide (MAPb$I_3$) perovskite devices. The two molecules exhibit a similarly high dipole moment that is further enhanced when the molecules are aggregated. Consequently, increasing the layer thickness results in a gradual increase in the $V_{OC}$ of the photovoltaic devices. While both derivatives are equally effective in hole-blocking due to their high ionisation potential, the different extent of conjugation leads to very different electron transport levels. Photovoltaic performance characterisation shows that these variations are mitigated by the increase in the built-in potential of the devices, resulting in an overall very similar increase in $V_{OC}$. To confirm the experimental results, the effect of the different HBL electron transport levels is investigated by numerical device simulations that demonstrate that the high dipole introduced by the HBL is sufficient to compensate for both the energetic mismatch at the PEDOT:PSS interface and the variation in the electron transport level of the HBL.

# 2. Results and Discussion

## 2.1 Synthesis of Electrolytes DTI-BF$_4$ and DDTI-BF$_4$

The nitrogen scaffolds serving as starting point of the synthesis were prepared according to a previously reported procedure[39]. 1.6 eq. and 6 eq. 2,3-Dichloro-5,6-dicyano-1,4-benzoquinone (DDQ) were used for dehydrogenation, respectively. Tetrafluoroboric acid (HBF$_4$) was added to obtain the final 2,3-diphenyl-[1,2,4]triazino[3,2-a]isoquinoline-5-ium tetrafluoroborate (DTI-BF$_4$) and 2,3-diphenyl-6,7-dihydro-[1,2,4]triazino[3,2-a]isoquinoline-5-ium tetrafluoroborate (DDTI-BF$_4$) as depicted in Scheme 1. Detailed description of the synthesis procedures, as well as characterization using standard methods can be found in the Supplementary Information (Scheme S1, Figures S1-S3).

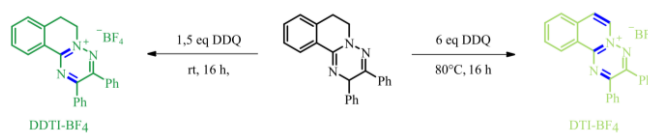

Scheme 1. Synthesis of TI-BF$_4$ compounds.

## 2.2 Effect of TI-BF$_4$ Aggregation on Photovoltaic Device Performance

To investigate the applicability of TI-BF$_4$ as hole-blocking layers in perovskite solar cells, they were incorporated in inverted architecture solar cells with the structure: glass/ITO/PEDOT:PSS/MAPbI$_3$/PCBM/HBL/Ag (**Figure 1a**). For this purpose, devices with different HBL thicknesses, achieved by varying the concentration of the HBL solution, were fabricated and compared to reference devices with BCP as HBL. We note that the morphology of both derivatives on PCBM is very similar as evidenced by atomic force microscopy measurements (Supplementary Information, Figure S4). **Figure 1b** summarises the photovoltaic performance for both TI-BF$_4$ derivatives. As expected, the use of BCP as HBL results in a relatively low V$_{OC}$ of only 0.96-0.98 V, in agreement with previous reports[40–42]. Replacing the BCP layer with either of the TI-BF$_4$ derivatives results in an increase in the V$_{OC}$. As the concentration of the TI-BF$_4$ is gradually increased from 0.25 mg/ml to 0.75 mg/ml, the V$_{OC}$ continuously increases reaching an average of 1.05-1.07 V, and a maximum of 1.09 V. The short-circuit current and fill factor remain largely unchanged when compared to the reference BCP device resulting in an overall enhancement in device power conversion efficiency by approximately 2%. Further increase of the thickness (corresponding to a solution concentration of 1 mg/ml) is not beneficial, since especially the fill factor of the devices suffers, most likely as a result of electron transport through the layer becoming a limiting factor.

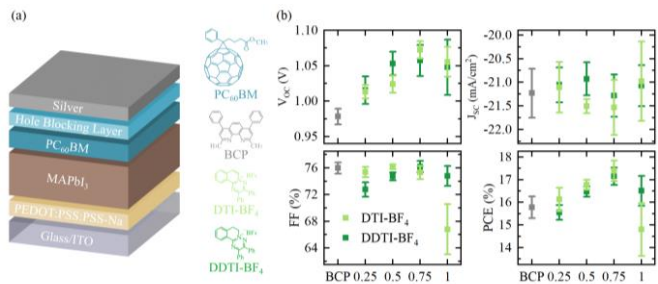

Figure 1. (a) Photovoltaic device structure with chemical structures of the materials used as extraction layers (b) Photovoltaic performance parameters of devices with increasing HBL thickness and reference BCP devices. Concentration ranges from 0.25 mg/ml to 1 mg/ml in methanol.

The increase in the V$_{OC}$ is unlike what has been previously observed for HBLs with different thickness[37]. This increase suggests that the dipole induced by the HBLs is increasing with increasing layer thickness, which can be triggered by aggregation of the TI-BF$_4$ molecules. To investigate what influence aggregation may have on the dipole formed by the layer we performed theoretical calculations on either isolated molecules or on molecules in their crystalline structure. **Figure 2a** depicts the electron density, which was plotted (isovalue = 0.01) and colorized according to a normalized gradient to highlight the differences in the electron density magnitude and the simulated dipoles of molecules DTI-BF$_4$ and DDTI-BF$_4$. **Figures 2b** and **2c** show the Löwdin charge population analysis of the atoms in the intersecting ring perpendicular to the dipole moment in the isolated and in the crystal structure of the molecules, respectively. In their isolated state both molecules DTI-BF$_4$ and DDTI-BF$_4$ show a similar dipole moment of 9.44 D and 9.73 D, respectively. The Löwdin charges of the atoms in the intersecting ring perpendicular to the dipole moment add to 0.083 and 0.094. As shown in **Figure 2c**, however, both molecules DTI-BF$_4$ and DDTI-BF$_4$ exhibit a compact aggregation behaviour leading to a considerably increased dipole-dipole interaction stemming from its inverse cubic distance dependence. Furthermore, an enhancement of the positive Löwdin charge of the two rings, intersecting the dipole moment vector, corroborates resulting polarization effects due to the compactness of the aggregation. Consequently, the total dipole moment of both aggregates is enhanced as compared to the molecules in the isolated states.

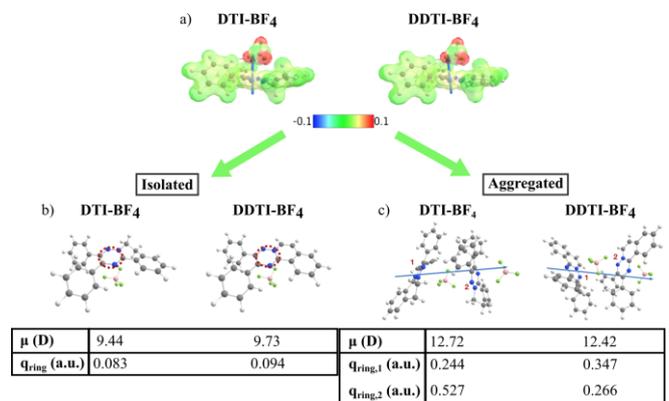

Figure 2. (a) Simulated electron density and dipole for molecules DTI-BF$_4$ and DDTI-BF$_4$. Löwdin charge population analysis of the atoms in the intersecting ring perpendicular to the dipole moment in the (b) isolated and (c) crystal structure state.

### 2.3 Effect of Electron Transport Levels on Photovoltaic Device Performance

To characterise the energy levels of DTI-BF$_4$ and DDTI-BF$_4$, ultra-violet photoemission spectroscopy (UPS) measurements were performed on glass/ITO/PCBM/DTI-BF$_4$,DDTI-BF$_4$ films yielding an ionization potential of 7 eV for both compounds (**Figure 3a**). For better comparison the weak PCBM peak still visible in the spectra of DTI-BF$_4$ and DDTI-BF$_4$ was subtracted in Figure 3a. The bandgap of both compounds in thin films was investigated using UV-vis spectroscopy (**Figure 3b**). Fitting the absorption onset of the spectra yields a bandgap of 2.6 eV and 2.9 eV for DTI-BF$_4$ and DDTI-BF$_4$, respectively. The complete energetic picture for both compounds as well as for PCBM and BCP is depicted in **Figure 3c**.

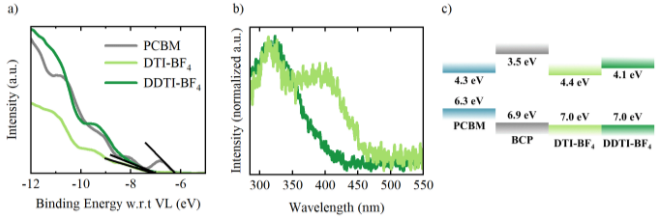

*Figure 3. (a) Ultra-violet photoemission spectroscopy measurements on PCBM, DTI-BF$_4$ and DDTI-BF$_4$ the black lines indicate the ionization potentials derived from the HOMO cutoff. (b) UV-vis spectra of DTI-BF$_4$ and DDTI-BF$_4$. (c) Summary of the energy levels of BCP, DTI-BF$_4$ and DDTI-BF$_4$ HBLs with PCBM as reference.*

The deep-lying highest occupied molecular orbital (HOMO) of both electrolytes suggests that their hole-blocking is equally effective as that of BCP. However, the different positions of the lowest unoccupied molecular orbital (LUMO) of the three HBLs with respect to the PCBM LUMO raises the question of the effect of the electron transport level on the performance of the solar cells. Although the LUMO of BCP lies significantly higher than the LUMO of PCBM, efficient electron transport via the BCP layer is possible due to the formation of gap states as a result of electrode evaporation.[43,44] In the case of the DTI-BF$_4$ and DDTI-BF$_4$ derivatives, their LUMOs are much deeper than that of BCP with DTI-BF$_4$ being slightly below and DDTI-BF$_4$ slightly above the LUMO of PCBM. These deviations from the PCBM LUMO may influence the device performance: for example, a lower lying LUMO may result in a lower built-in potential in the device, while alternatively a higher lying LUMO may result in an extraction barrier of electrons from the PCBM layer. To compare the photovoltaic performance of devices with both derivatives, devices with HBL layers deposited from the optimal concentration of 0.75 mg/ml were fabricated and characterised. **Figure 4a** shows the J-V curves of champion devices fabricated with either BCP or DTI-BF$_4$ and DDTI-BF$_4$ HBLs and **Table 1** summarising their photovoltaic parameters. The corresponding external quantum efficiency (EQE) spectra are shown in **Figure 4b**. Statistics acquired from 30 devices are shown in **Figure 4c**.

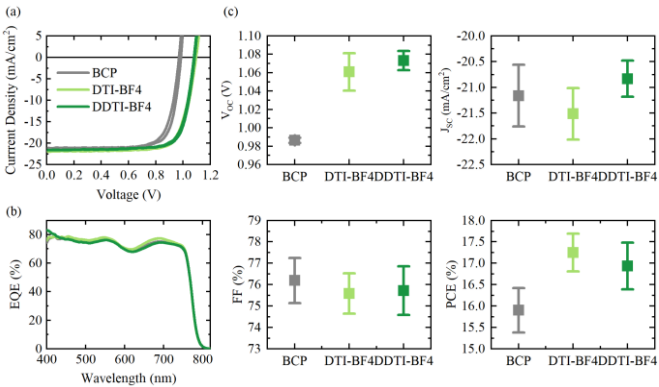

Figure 4. a) J-V-curves for champion devices with each HBL measured under one sun illumination d) External quantum efficiency spectra for each type of device. (c) Photovoltaic parameters statistics of 30 photovoltaic devices.

Despite the differences in the LUMO position between the two derivatives, they reach very similar photovoltaic performance. No significant difference in the built-in potential of the device can be observed from the dark current J-V characteristics (Supplementary Information, Figure S5), suggesting that even if variation in the electron transport level reduced the built-in potential of the device, the high dipole moment introduced by the HBL compensates for this non-ideality. Overall, with no variation in the fill factor and short-circuit current between the well-established BCP and the novel electrolyte derivatives, it is the difference in $V_{OC}$ which determines the overall improvement in device performance. On average the $V_{OC}$ increases by approximately 80 mV, with a maximum increase of 120 mV. Correspondingly, the maximum PCE is increased from 16.28 % for a BCP device to 18.33 % and 18.38 % for devices with DTI-BF$_4$ and DDTI-BF$_4$, respectively.

Table 1. Photovoltaic parameters of optimal MAPbI$_3$ devices with different hole-blocking layers. Forward Scan (from $J_{SC}$ to $V_{OC}$) and Reverse scan (from $V_{OC}$ to $J_{SC}$) parameters are provided.

|  | Forward Scan | | | | Reverse Scan | | | |
|---|---|---|---|---|---|---|---|---|
|  | Voc [V] | $J_{SC}$ [mA/cm$^2$] | FF [%] | PCE [%] | Voc [V] | $J_{SC}$ [mA/cm$^2$] | FF [%] | PCE [%] |
| **BCP** | 0.97 | -21.06 | 76.70 | 15.68 | 0.99 | -21.06 | 78.47 | 16.28 |
| **DTI-BF$_4$** | 1.09 | -21.98 | 76.10 | 18.19 | 1.10 | -21.98 | 76.08 | 18.33 |
| **DDTI-BF$_4$** | 1.08 | -21.57 | 77.27 | 18.00 | 1.09 | -21.57 | 78.51 | 18.38 |

To confirm that the electron transport level plays only a minor role in determining the device performance, we employed numerical device simulations in which we compare the simulated J-V characteristics of devices with HBLs with three different LUMOs. First, to simulate the case of BCP, we introduce a HBL with no dipole and an electron transport level at the same level as PCBM. Next, we compare the two electrolyte HBLs which introduce a dipole of the same strength, yet exhibiting different electron transport levels: DTI-BF$_4$ at 0.1 eV below and DDTI-BF$_4$ at 0.2 eV above the LUMO of PCBM. As a reference a HBL of the same dipole and an electron transport level equivalent to that of PCBM is also simulated (Figure S6). The results of the numerical simulations (**Figure 5**) confirm that these variations of the electron transport level have only a marginal effect on the $V_{OC}$ with all three dipolar cases resulting in a similar $V_{OC}$, which is significantly higher than that of the non-dipolar HBL BCP.

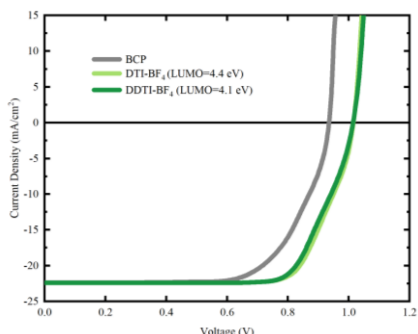

Figure 5. Simulated J-V curves of reference BCP hole-blocking layer and the two dipoles with different LUMO energies.

*3. Conclusion*

To summarise, we report on the synthesis and application of two novel electrolyte derivatives in inverted architecture perovskite photovoltaic devices. We demonstrate that aggregation plays an important role in determining the induced dipole and the final open-circuit voltage of the devices. We also show that while the molecules exhibit electron transport levels that vary by 0.3 eV, this variation has little to no effect on the final device performance. Our study demonstrates that dipolar HBLs are a powerful method to enhance the performance of perovskite photovoltaic devices, with particular emphasis on the importance of the dipole strength induced by the layer, rather than the position of charge transport levels.

**4. Experimental Methods**

**Materials:** Methylammonium iodide ($CH_3NH_3I$) was purchased from GreatCell Solar. PEDOT:PSS was purchased from Heraeus Deutschland GmbH&Co and PCBM (99.5%) was purchased from Solenne BV. All other materials were purchased from Sigma-Aldrich and used as received.

**Photovoltaic Device Fabrication:** Pre-patterned indium tin oxide (ITO) coated glass substrates (PsiOTech Ltd., 15 Ohm/sq) were subsequently cleaned in 2 % hellmanex detergent, deionized water, acetone and isopropanol in an ultra-sonicate bath. Shortly before spin coating the first layer, the samples were treated for 8 min with an oxygen plasma. PEDOT:PSS prepared based on a previous report [11] was spin coated on the substrates with 4000 rpm 30 s and annealed at 150 °C for 15 min. The samples were then transferred into a dry air filled glovebox (RH < 0.5 %). A lead acetate trihydrate $MAPbI_3$ recipe [45] was used for forming the $MAPbI_3$ perovskite layer, in detail the perovskite solution was spin coated at 2000 rpm for 60 s. After blowing 25 s with a dry air gun and resting for 5 min, the films were annealed at 100 °C for another 5 min. All samples were then transformed into a nitrogen filled glovebox where a PCBM (20 mg/ml in chlorobenzene) layer was dynamically spin coated at 2000 rpm for 30 s on the perovskite films. The films were annealed for 10 min at 100 °C. Finally, either BCP (0.5 mg/ml in isopropanol) or the $TI-BF_4$ molecules (various concentrations in methanol) were dynamically spin coated at 4000 rpm for 30 s. The devices were completed with an 80 nm silver electrode that was deposited via thermal evaporation under high vacuum.

**Photovoltaic Device Characterization:** The current density-voltage (J-V) was measured by a computer controlled Keithley 2450 Source Measure Unit under simulated AM 1.5G sunlight with 100 mW cm$^{-2}$ irradiation (Abet Sun 3000 Class AAA solar simulator). The light intensity was calibrated with a Si reference cell (NIST traceable, VLSI) and corrected by measuring the spectral mismatch between the solar spectrum, the spectral response of the perovskite solar cell and the reference cell. The cells were scanned from forward bias to short circuit and reverse at a rate of 0.25 V s$^{-1}$.

**UV-vis absorption:** The $TI-BF_4$ molecules were dissolved at 0.5 mg/ml in methanol and spin coated on clean glass substrates. The UV-Vis spectra of the resulting thin films were measured at room temperature with a Jasco UV-Vis V670 spectrometer.

**Ultra-violet Photoemission Spectroscopy (UPS):** Ultra-violet photoemission spectroscopy measurements were performed on PCBM/BCP and PCBM/$TI-BF_4$ films to characterize their ionization potential. The samples were transferred to an ultrahigh vacuum (UHV) chamber of the PES system (Thermo Scientific ESCALAB 250Xi) for measurements. UPS measurements were carried out using a double-differentially pumped He discharge lamp (hν = 21.22 eV) with a pass energy of 2 eV and a bias of -10 V.

**Dipole Simulation:** Geometry optimization of the monomers were done with pbe0/pc-1 and the D3 dispersion correction within the TeraChem software package.[1,2,3] For the single point calculations the double-hybrid functional PWPB95 with the D3 dispersion correction including Becke-Johnson damping and the def2-TZVPD basis set was used as implemented in the ORCA software package.[4]

**Numerical Device Simulations:** A previously developed model that includes the contributions of charges and ions has been used to simulate the influence of incorporating $TI-BF_4$ HBL into the device structure.[46]

**Acknowledgements**

This work has received financial support of the Deutsche Forschungsgemeinschaft (DFG) SFB 1249 projects A03, B01 and C04. This project has also received funding from the European Research Council (ERC) under the European Union's Horizon 2020 research and innovation program (ERC Grant Agreement n° 714067, ENERGYMAPS). N.T. acknowledges support by the Israel Science Foundation (grant no. 488/16), the Adelis Foundation for renewable energy research within the framework of the Grand Technion Energy